\documentclass[twocolumn,a4paper,nofootinbib,showpacs,aps,floatfix]{revtex4}
\usepackage{graphicx}
\usepackage{bm}
\usepackage{amsmath}
\usepackage{color,colordvi} 
\newcommand{\ignore}[1]{}
\def\beq{\begin{equation}}
\def\eeq{\end{equation}}

\def\pr{\prime} 
\begin{document}
\title{1D Photonic Crystals with a Sawtooth Refractive Index} 
\author{G. V. Morozov}
\affiliation{Scottish Universities Physics Alliance (SUPA),
Thin Film Centre, University of the West of Scotland, Paisley PA1 2BE, Scotland, UK
}
\author{D. W. L. Sprung}
\affiliation{
  Department of Physics and Astronomy, McMaster University\\
  Hamilton, Ontario L8S 4M1 Canada
}
\author{J. Martorell}
\affiliation {Dept.
  d'Estructura i Constituents de la Materia, Facultat F\'{\i}sica,\\
   University of Barcelona, Barcelona 08028, Spain
}

\date{February 1, 2013}

\begin{abstract}
Exact analytical results (in terms of Bessel functions) for the bandgaps, 
reflectance, and transmittance of one-dimensional photonic crystals 
with a sawtooth refractive index profile on the period
are derived for the first time. 
This extends a group of exactly solvable models 
of periodic refractive indices. 
The asymptotic approximations of the above exact results 
have been also obtained.
\end{abstract}
\pacs{
42.70.Qs,       
42.25.Bs,       
78.67.-n       
}
\maketitle
\section{Introduction}
Recently there has been a renewal of interest in the properties of 1D photonic
crystals based on graded index slabs. Rauh {\it et al}. \cite{Ra10} have considered in some
detail optical properties of periodic systems consisting of slabs whose permittivities {$n^2(z)$ 
increase linearly with depth $z$}. They also considered periodic systems in which 
alternating layers have quadratically increasing and decreasing
permittivities. Their work pointed out the relevance of such 1D
photonic crystals for technological applications. They stressed in particular
that to a good approximation all the bandgaps have
the same width. 

In this paper we consider the properties of 1D photonic crystal 
constructed of layers with linearly increasing refractive index {$n(z)$}, 
giving a quadratic increase of the permittivity {$n^2(z)$}.  
To be precise, for a slab extending from $z=0$ to $z=d$ we take 
\begin{equation}                        
n(z) = n_a + (n_b-n_a)\frac{z}{d} \equiv \frac{n_b-n_a}{d}(z-z_0),
\label{eq:tp14c}
\end{equation}
with $z_0 \equiv -  n_a d /(n_b-n_a)$. As seen in Fig.~1 the refractive
index is periodic, increasing linearly from $n_a$ to $n_b$ inside each
layer, and falls sharply at the cell boundaries.  
\begin{figure} [htb]                         
\includegraphics[width=8cm]{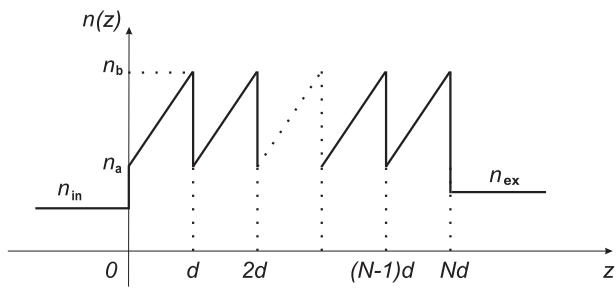}
\caption{1D photonic crystal with a sawtooth refractive index.}
\label{Fig1saw.eps}
\end{figure}
From data in Adachi \cite{Adachi} one finds that the
permittivity of the ternary alloy Al$_x$Ga$_{1-x}$As 
is approximately linear in the relative content $x >
0.4$ of Al for wavelengths 10000 \AA \, $ > \lambda > $ 6000 \AA, so 
photonic crystals with linearly graded refractive index layers are within
present technological possibilities.
 
We will express the dispersion equation for the bandgaps
of these photonic crystals as well as their reflectance and transmittance
in terms of Bessel functions of fractional order. 
This adds to a group of exactly solvable models 
of periodic refractive indices which is restricted at the moment
to periodic structures with a step-layered profile of the index on the period
(binary and ternary photonic crystals) and by some specific two-parameter
sinusoidal periodic potentials \cite{Wu,Urumov}. 
By  means of asymptotic expansions we derive approximate sinusoidal 
expressions showing that the bandgaps of our sawtooth periodic potential 
tend to a constant width as the wavelength decreases.
For simplicity, as in Rauh's work \cite{Ra10}, 
we restrict our attention to normal propagation.

\section{Exact Results}

Description of optical wave propagation of linearly polarized light
of circular frequency $\omega$ through a periodic structure, 
with real refractive index $n(z+d)=n(z)$,
in case of normal incidence, reduces from Maxwell's equations
to the Helmholtz equation: 
\begin{equation}                                
E_{zz}''+k_0^2 n^2(z) E(z) =0,
\end{equation}
where the total electric and magnetic fields of propagating light are expressed as
\begin{equation}                                
\begin{split}
{\bf E} & = E(z)\exp(-i\omega t)\,\hat{{\bf y}},\\
{\bf B} & = \frac{i}{k_0}\frac{dE(z)}{dz}\exp(-i\omega t)\,\hat{{\bf x}},
\end{split}
\end{equation}
Applying the transfer matrix method for periodic potentials \cite{YY}, 
the field $E(z)$ and its derivative $dE(z)/dz$ 
at the edge points of the system $z=0$ and $z=Nd$ are related by
\begin{equation}                                
\left[ {\begin{array}{*{20}c}
   {E\left( {Nd} \right)}  \\
   {E'\left( {Nd} \right)}  \\
\end{array}} \right] = \left[ {\begin{array}{*{20}c}
   {v'\left( d \right)} & { - v\left( d \right)}  \\
   { - u'\left( d \right)} & {u(d)}  \\
\end{array}} \right]^{\,N} \left[ {\begin{array}{*{20}c}
   {E\left( 0 \right)}  \\
   {E'\left( 0 \right)}  \\
\end{array}} \right],
\end{equation}
where $u(z)$ and $v(z)$ are the so-called normalized solutions,
which satisfy the boundary conditions
\begin{equation}                                
\left[ {\begin{array}{*{20}c}
   {u\left( 0 \right)}  \\
   {u'\left( 0 \right)}  \\
\end{array}} \right] = \left[ {\begin{array}{*{20}c}
   1  \\
   0  \\
\end{array}} \right], \quad \left[ {\begin{array}{*{20}c}
   {v\left( 0 \right)}  \\
   {v'\left( 0 \right)}  \\
\end{array}} \right] = \left[ {\begin{array}{*{20}c}
   0  \\
   1  \\
\end{array}} \right].
\end{equation}
The $N'$th power $W^N$ of a unimodular $W$-matrix as occurs in Eq.~(4)
can be expressed in terms of $W$ and the unit matrix $[1]$ 
as shown in Ref.~\cite{FPP}
\begin{equation}                                
W^{N}= \frac{\sin\,N\phi}{\sin\phi}\,\,\,W
-\frac{\sin(N-1)\phi}{\sin\phi}\,\,\,[1]\,,
\end{equation} 
where the dispersion equation for the Bloch phase $\phi$ takes the form
\begin{equation}                                
\cos\phi \equiv \frac{1}{2}\,{\rm Tr}\,W = \frac{1}{2}\left[u(d)+v'(d)\right]. 
\end{equation}  
The advantage of the transfer matrix method is that once we have solved 
for $W$ of a single layer, the result for 
$N$ layers is trivial, by use of Eq.~(6). 

In dimensionless units,  
Eq.~(2) with the refractive index defined by Eq.~(1),
takes the form
\begin{equation}                                
\frac{d^2E(\tilde{z})}{d\tilde{z}^2} + \frac{1}{4}\tilde{z}^2 E(\tilde{z}) = 0,
\label{eq:tp16}
\end{equation} 
where the dimensionless variable $\tilde{z}$ is
\begin{equation}                                
\tilde{z} = \sqrt{\gamma}\,(z-z_0), \quad
\gamma \equiv \frac{2 k_0 (n_b-n_a)}{d}\,.
\end{equation}
It is straightforward to prove 
(see Appendix A) that the normalized solutions of Eq.~(8) are Bessel functions of order $\nu = \pm 1/4$, 
multiplied by $\tilde{z}^{1/2}$\,, i.e. 
\begin{equation}                                
\begin{split}                                   
\tilde{u}(\tilde{z})& = {\frac{\pi}{2^{1/4}\Gamma(1/4)}}\,\tilde{z}^{1/2} 
J_{-1/4}\left(\frac{\tilde{z}^2}{4}\right) \sim 1 - \frac{\tilde{z}^4}{48} \cdots, \\ 
\tilde{v}(\tilde{z})& = \frac{\pi}{2^{3/4}\Gamma(3/4)}\,\tilde{z}^{1/2}
J_{1/4}\left(\frac{\tilde{z}^2}{4}\right) 
\sim \tilde{z} - \frac{\tilde{z}^5}{80} \cdots .
\label{eq:ex5}
\end{split}
\end{equation}
Using standard relations among  Bessel functions, see chapter 9 in \cite{AS}, 
one finds  
\begin{equation}                                
\begin{split}
\frac{d\tilde{u}(\tilde{z})}{d\tilde{z}}& = -\frac{\pi}{2^{5/4}\Gamma(1/4)} \tilde{z}^{3/2} J_{3/4}\left(\frac{\tilde{z}^2}{4}\right),\\ 
\frac{d\tilde{v}(\tilde{z})}{d\tilde{z}}& = \frac{\pi}{2^{7/4}\Gamma(3/4)} \tilde{z}^{3/2} J_{-3/4}\left(\frac{\tilde{z}^2}{4}\right). 
\label{eq:ex9}
\end{split}
\end{equation}

If  we now change the variable $\tilde{z}$ in Eq.~(10)
back to the original variable $z$,
we obtain another set of the fundamental solutions, 
$\tilde{u}(z)$ and $\tilde{v}(z)$, of Eq.~(2)
\begin{equation}
\begin{split}                                   
\tilde{u}(z)& = {\frac{\pi}{2^{1/4} \Gamma(1/4)}}\,\left[\sqrt\gamma(z-z_0)\right]^{1/2}J_{-1/4}\left[\frac{\gamma(z-z_0)^2}{4}\right], \\ 
\tilde{v}(z)& = \frac{\pi}{2^{3/4}\Gamma(3/4)}\,\left[\sqrt\gamma(z-z_0)\right]^{1/2}
J_{1/4}\left[\frac{\gamma(z-z_0)^2}{4}\right].
\end{split}
\end{equation}

To construct the matrix $W$ of Eq.~(4), 
one has to express $u(z)$ and $v(z)$ 
in terms of $\tilde{u}(z)$ and $\tilde{v}(z)$.
Taking into account Eq.~(5), one obtains
\begin{equation}                                
\begin{split}
u(z) & =\frac{\tilde{v}'(0)}{\tilde{w}(0)}\,\tilde{u}(z)
-\frac{\tilde{u}'(0)}{\tilde{w}(0)}\,\tilde{v}(z),\\
v(z) & =-\frac{\tilde{v}(0)}{\tilde{w}(0)}\,\tilde{u}(z)
+\frac{\tilde{u}(0)}{\tilde{w}(0)}\,\tilde{v}(z),
\end{split}
\end{equation}
where the Wronskian $\tilde{w}(z)$ of the fundamental solutions  $\tilde{u}(z)$
and $\tilde{v}(z)$ of Eq.~(2) at the point $z=0$ is
\begin{equation}                                
\tilde{w}(0) = \tilde{u}(0)\tilde{v}'(0)-\tilde{v}(0)\tilde{u}'(0)=\sqrt\gamma. 
\end{equation}
With the aid of newly introduced  dimensionless parameters, 
\begin{equation}                                
\begin{split}
z_a = -\sqrt\gamma\,z_0 = \left(\frac{2k_0d\,n_a^2}{n_b-n_a}\right)^{1/2},\\
z_b = \sqrt\gamma\,(d-z_0) =  \left(\frac{2k_0d\,n_b^2}{n_b-n_a}\right)^{1/2},
\end{split}
\label{eq:ex2}
\end{equation}
the values of $\tilde{u}(0)$, $\tilde{v}(0)$, $\tilde{u}'(0)$, $\tilde{v}'(0)$
take the form
\begin{equation}                                
\begin{split}
\tilde{u}(0)& = \frac{\pi}{2^{1/4}\Gamma(1/4)}z_a^{1/2}
J_{-1/4}\left(\frac{z_a^{2}}{4}\right),\\
\tilde{v}(0)& = \frac{\pi}{2^{3/4}\Gamma(3/4)}z_a^{1/2}
J_{1/4}\left(\frac{z_a^{2}}{4}\right),\\
\tilde{u}'(0) \equiv \frac{d\tilde{u}(z)}{dz}\bigg{|}_{z=0}
& = -\sqrt\gamma\,\,\frac{\pi}{2^{5/4}\Gamma(1/4)}z_a^{3/2}
J_{3/4}\left(\frac{z_a^2}{4}\right),\\
\tilde{v}'(0) \equiv \frac{d\tilde{v}(z)}{dz}\bigg{|}_{z=0}
& = \sqrt\gamma\,\,\frac{\pi}{2^{7/4}\Gamma(3/4)}z_a^{3/2}
J_{-3/4}\left(\frac{z_a^{2}}{4}\right)
\end{split}
\end{equation}
while at $z=d$,  $\,\,\,$  
$\tilde{u}(d)$, $\tilde{v}(d)$, $\tilde{u}'(d)$, $\tilde{v}'(d)$ are given by
\begin{equation}                                
\begin{split}
\tilde{u}(d)& 
= \frac{\pi}{2^{1/4}\Gamma(1/4)}z_b^{1/2}
J_{-1/4}\left(\frac{z_b^{2}}{4}\right),\\
\tilde{v}(d)& = \frac{\pi}{2^{3/4}\Gamma(3/4)}z_b^{1/2}
J_{1/4}\left(\frac{z_b^{2}}{4}\right),\\
\tilde{u}'(d) \equiv \frac{d\tilde{u}(z)}{dz}\bigg{|}_{z=d}
& = -\sqrt\gamma\,\,\frac{\pi}{2^{5/4}\Gamma(1/4)}z_b^{3/2}
J_{3/4}\left(\frac{z_b^{2}}{4}\right),\\
\tilde{v}'(d) \equiv \frac{d\tilde{v}(z)}{dz}\bigg{|}_{z=d}
& = \sqrt\gamma\,\,\frac{\pi}{2^{7/4}\Gamma(3/4)}z_b^{3/2}
J_{-3/4}\left(\frac{z_b^{2}}{4}\right)~.
\end{split}
\end{equation}

\begin{widetext}
The elements of the $W$-matrix are then
\begin{equation}                                
\begin{split}
W_{11} & = v'(d) 
= \frac{\pi}{2^{5/2}}\,\,\sqrt{\frac{n_b}{n_a}}\,\,z_az_b\,
\bigg[J_{1/4}\left(z_a^2/4\right)\,J_{3/4}\left(z_b^2/4\right)
+J_{-1/4}\left(z_a^2/4\right)\,J_{-3/4}\left(z_b^2/4\right)\bigg],\\
\frac{W_{21}}{k_0} & = - \frac{u'(d)}{k_0}
= \frac{\pi}{2^{5/2}}\,\,\sqrt{n_an_b}\,\,z_az_b\,
\bigg[J_{-3/4}\left(z_a^2/4\right)\,J_{3/4}\left(z_b^2/4\right)
-J_{3/4}\left(z_a^2/4\right)\,J_{-3/4}\left(z_b^2/4\right)\bigg],\\
k_0W_{12} & = - k_0\,v(d)  
= \frac{\pi}{2^{5/2}}\,\,\sqrt{\frac{1}{n_an_b}}\,\,z_az_b\,
\bigg[J_{1/4}\left(z_a^2/4\right)\,J_{-1/4}\left(z_b^2/4\right)
-J_{-1/4}\left(z_a^2/4\right)\,J_{1/4}\left(z_b^2/4\right)\bigg],\\
W_{22} & = u(d) 
= \frac{\pi}{2^{5/2}}\,\,\sqrt{\frac{n_a}{n_b}}\,\,z_az_b\,
\bigg[J_{-3/4}\left(z_a^2/4\right)\,J_{-1/4}\left(z_b^2/4\right)
+J_{3/4}\left(z_a^2/4\right)\,J_{1/4}\left(z_b^2/4\right)\bigg],
\end{split}
\end{equation}
while the dispersion equation (7) takes the form  
\begin{equation}                        
\begin{split}
\cos \phi = \frac{\pi}{2^{7/2}}\,\,z_a z_b
& \bigg[\sqrt{\frac{n_a}{n_b}}\, J_{-3/4}\left(z_a^2/4\right)\,J_{-1/4}\left(z_b^2/4\right)
+ \sqrt{\frac{n_a}{n_b}}\,J_{3/4}\left(z_a^2/4\right)\,J_{1/4}\left(z_b^2/4\right)\\
& + \sqrt{\frac{n_b}{n_a}}\,J_{1/4}\left(z_a^2/4\right)\,J_{3/4}\left(z_b^2/4\right)
+ \sqrt{\frac{n_b}{n_a}}\,J_{-1/4}\left(z_a^2/4\right)\,J_{-3/4}\left(z_b^2/4\right)\bigg]. 
\end{split} 
\label{eq:ba5}
\end{equation}
Eq.~\ref{eq:ba5} is the exact dispersion relation for our  sawtooth 1D photonic crystal. 
The amplitude reflection $r$ and transmission $t$ coefficients are
\begin{equation}                        
\begin{split}
r = & \frac{W^N_{11}-\displaystyle\frac{n_{\rm ex}}{n_{\rm in}}W^N_{22}
+i\left[k_0n_{\rm ex}\,W^N_{12}+\frac{1}{k_0n_{\rm in}}W^N_{21}\right]}
{W^N_{11}+\displaystyle\frac{n_{\rm ex}}{n_{\rm in}}W^N_{22}+i\left[k_0n_{\rm ex}W^N_{12}-\frac{1}{k_0n_{\rm in}}W^N_{21}\right]}\,,\\
t = & \frac{2}{W^N_{11}+\displaystyle\frac{n_{\rm ex}}{n_{\rm in}}W^N_{22}+i\left[k_0n_{\rm ex}\,W^N_{12} -\frac{1}{k_0n_{\rm in}}W^N_{21}\right]}\,,
\end{split}
\end{equation}
where the elements of the $W$-matrix are given in exact form by Eq.~(18)
and the elements of the $W^N$-matrix are obtained by use of Eq.~(6).
The indices of refraction in the incident and exit media are $n_{\rm in}$ and
$n_{\rm ex}$ as seen in Fig.~1.

\section{Asymptotic Estimates}

The motivation for exploring those is to have  simple analytic approximations for the Bloch phase $\cos \phi$, 
as well as for the reflection and transmission coefficients.
Using Hankel's asymptotic formula,   section 9.2 of Ref. \cite{AS}, 
one obtains the elements of the $W$ matrix in the form
\begin{equation}                                
\begin{split}
W_{11} & = \sqrt{\frac{n_b}{n_a}}\,\bigg[P_{1/4}(z_a^2/4)\,P_{3/4}(z_b^2/4) + Q_{1/4}(z_a^2/4)\,Q_{3/4}(z_b^2/4)\bigg]\cos\left(z_a^2/4-z_b^2/4\right) \\
& + \sqrt{\frac{n_b}{n_a}}\,\bigg[P_{1/4}(z_a^2/4)\,Q_{3/4}(z_b^2/4) - Q_{1/4}(z_a^2/4)\,P_{3/4}(z_b^2/4)\bigg]\sin\left(z_a^2/4-z_b^2/4\right), \\
\frac{W_{21}}{k_0} & = -\sqrt{n_a\,n_b}\,\bigg[P_{3/4}(z_a^2/4)\,P_{3/4}(z_b^2/4) +  Q_{3/4}(z_a^2/4)\,Q_{3/4}(z_b^2/4)\bigg]\sin\left(z_a^2/4-z_b^2/4\right) \\
& + \sqrt{n_a\,n_b}\,\bigg[P_{3/4}(z_a^2/4)\,Q_{3/4}(z_b^2/4) - Q_{3/4}(z_a^2/4)\,P_{3/4}(z_b^2/4)\bigg]\cos\left(z_a^2/4-z_b^2/4\right), \\
k_0W_{12} & = \frac{1}{\sqrt{n_a\,n_b}}\,
\bigg[P_{1/4}(z_a^2/4)\,P_{1/4}(z_b^2/4) +  Q_{1/4}(z_a^2/4)\,Q_{1/4}(z_b^2/4)\bigg]\sin\left(z_a^2/4-z_b^2/4\right) \\
& + \frac{1}{\sqrt{n_a\,n_b}}\,\bigg[P_{1/4}(z_b^2/4)\,Q_{1/4}(z_a^2/4) - P_{1/4}(z_a^2/4)\,Q_{1/4}(z_b^2/4)\bigg]\cos\left(z_a^2/4-z_b^2/4\right),\\
W_{22} & = \sqrt{\frac{n_a}{n_b}}\,\bigg[P_{1/4}(z_b^2/4)\,P_{3/4}(z_a^2/4) + Q_{1/4}(z_b^2/4)\,Q_{3/4}(z_a^2/4)\bigg]\cos\left(z_a^2/4-z_b^2/4\right) \\
& + \sqrt{\frac{n_a}{n_b}}\,\bigg[P_{3/4}(z_a^2/4)\,Q_{1/4}(z_b^2/4) - P_{1/4}(z_b^2/4)\,Q_{3/4}(z_a^2/4)\bigg]\sin\left(z_a^2/4-z_b^2/4\right),
\end{split}
\end{equation}
where
\begin{equation}                                
\begin{split}                                           
P_{\nu}(z) & = 1 -\frac{(4\nu^2-1)(4\nu^2-9)}{2!(8z)^2} 
- \frac{(4\nu^2-1)(4\nu^2-9)(4\nu^2-25)(4\nu^2-49)}{4!(8z)^4} \ldots \,,\\ 
Q_{\nu}(z) & = \frac{4\nu^2-1}{8z} -\frac{(4\nu^2-1)(4\nu^2-9)(4\nu^2-25)}{3!(8z)^3} \ldots \,,
\label{eq:asi1}
\end{split}
\end{equation}
and the dispersion equation in the form
\begin{equation}                                
\begin{split}
\cos\,\phi & = \frac{1}{2}\sqrt{\frac{n_b}{n_a}}\,\bigg[P_{1/4}(z_a^2/4)\,P_{3/4}(z_b^2/4) 
+ Q_{1/4}(z_a^2/4)\,Q_{3/4}(z_b^2/4)\bigg]\cos\left(z_a^2/4-z_b^2/4\right) \\
& +\frac{1}{2}\sqrt{\frac{n_a}{n_b}}\,\bigg[P_{1/4}(z_b^2/4)\,P_{3/4}(z_a^2/4) 
+ Q_{1/4}(z_b^2/4)\,Q_{3/4}(z_a^2/4)\bigg]\cos\left(z_a^2/4-z_b^2/4\right) \\
& + \frac{1}{2} \sqrt{\frac{n_b}{n_a}}\,\bigg[P_{1/4}(z_a^2/4)\,Q_{3/4}(z_b^2/4) - Q_{1/4}(z_a^2/4)\,P_{3/4}(z_b^2/4)\bigg]\sin\left(z_a^2/4-z_b^2/4\right)\\
& + \frac{1}{2} \sqrt{\frac{n_a}{n_b}}\,\bigg[P_{3/4}(z_a^2/4)\,Q_{1/4}(z_b^2/4) - P_{1/4}(z_b^2/4)\,Q_{3/4}(z_a^2/4)\bigg]\sin\left(z_a^2/4-z_b^2/4\right).
\end{split}
\end{equation}
Eqs.~(21)~-~(23) are the expressions we sought. 
By truncating the terms $P_\nu$ and $Q_\nu$  various asymptotic approximations are obtained. 
In particular, after some trigonometric simplifications,
the matrix elements  become in the first approximation ($P=1$, $Q = 0$) 
\begin{equation}                                
\begin{split}
W_{11} & = \sqrt{\frac{n_b}{n_a}}\,\cos\left(z_a^2/4-z_b^2/4\right),\\
\frac{W_{21}}{k_0} & = -\sqrt{n_a\,n_b}\,\sin\left(z_a^2/4-z_b^2/4\right),\\
k_0W_{12} & = \frac{1}{\sqrt{n_a\,n_b}}\,\sin\left(z_a^2/4-z_b^2/4\right),\\
W_{22} & = \sqrt{\frac{n_a}{n_b}}\,\cos\left(z_a^2/4-z_b^2/4\right),
\end{split}
\end{equation}
in the second ($P = 1$, $Q = (\nu^2 - 0.25)/2z$ ),  
\begin{equation}                                
\begin{split}
W_{11} & = \sqrt{\frac{n_b}{n_a}}\,\bigg[\cos\left(z_a^2/4-z_b^2/4\right)
+\bigg(\frac{3}{8z_a^2}+\frac{5}{8z_b^2}\bigg)\sin\left(z_a^2/4-z_b^2/4\right)\bigg],\\
\frac{W_{21}}{k_0} & = -\sqrt{n_a\,n_b}\,\bigg[\sin\left(z_a^2/4-z_b^2/4\right)
+\bigg(\frac{5}{8z_a^2}-\frac{5}{8z_b^2}\bigg)\cos\left(z_a^2/4-z_b^2/4\right)\bigg],\\
k_0W_{12} & = \frac{1}{\sqrt{n_a\,n_b}}\,\bigg[\sin\left(z_a^2/4-z_b^2/4\right)
-\bigg(\frac{3}{8z_a^2}-\frac{3}{8z_b^2}\bigg)\cos\left(z_a^2/4-z_b^2/4\right)\bigg],\\
W_{22} & = \sqrt{\frac{n_a}{n_b}}\,\bigg[\cos\left(z_a^2/4-z_b^2/4\right)
-\bigg(\frac{5}{8z_a^2}+\frac{3}{8z_b^2}\bigg)\sin\left(z_a^2/4-z_b^2/4\right)\bigg],
\end{split}
\end{equation}
while in the third,  
\begin{equation}                                
\begin{split}
W_{11} & = \sqrt{\frac{n_b}{n_a}}\,\bigg[\bigg(1-\frac{105}{128z_a^4}+\frac{135}{128z_b^4}-\frac{15}{64z_a^2z_b^2}\bigg)
\cos\left(z_a^2/4-z_b^2/4\right)
+\bigg(\frac{3}{8z_a^2}+\frac{5}{8z_b^2}\bigg)\sin\left(z_a^2/4-z_b^2/4\right)\bigg],\\
\frac{W_{21}}{k_0} & = -\sqrt{n_a\,n_b}\,\bigg[
\bigg(1+\frac{135}{128z_a^4}+\frac{135}{128z_b^4}+\frac{25}{64z_a^2z_b^2}\bigg)\sin\left(z_a^2/4-z_b^2/4\right)
+\bigg(\frac{5}{8z_a^2}-\frac{5}{8z_b^2}\bigg)\cos\left(z_a^2/4-z_b^2/4\right)\bigg],\\
k_0W_{12} & = \frac{1}{\sqrt{n_a\,n_b}}\,\bigg[
\bigg(1-\frac{105}{128z_a^4}-\frac{105}{128z_b^4}+\frac{9}{64z_a^2z_b^2}\bigg)\sin\left(z_a^2/4-z_b^2/4\right)
-\bigg(\frac{3}{8z_a^2}-\frac{3}{8z_b^2}\bigg)\cos\left(z_a^2/4-z_b^2/4\right)\bigg],\\
W_{22} & = \sqrt{\frac{n_a}{n_b}}\,\bigg[\bigg(1+\frac{135}{128z_a^4}-\frac{105}{128z_b^4}-\frac{15}{64z_a^2z_b^2}\bigg)
\cos\left(z_a^2/4-z_b^2/4\right)
-\bigg(\frac{5}{8z_a^2}+\frac{3}{8z_b^2}\bigg)\sin\left(z_a^2/4-z_b^2/4\right)\bigg].
\end{split}
\end{equation}
As one can see, the dispersion equation $\cos\phi=\frac{1}{2}(W_{11}+W_{22})$ takes a particularly simple form
in the first approximation
\begin{equation}                                
\cos\,\phi_1 = \frac{1}{2}\bigg(\sqrt{\frac{n_a}{n_b}}+\sqrt{\frac{n_b}{n_a}}\bigg)\,\cos\left(z_a^2/4-z_b^2/4\right).
\end{equation}
\end{widetext}

Figs.~2,~3,~4 illustrate the above approximations. 
In the second and third approximations, $\cos \phi$ of Eq.~(23) takes the form 
\begin{equation}				
\cos \phi_{\,2,3}=A_{2,3} \cos (z_a^2/4-z_b^2/4 + \delta_{\,2,3}),
\end{equation}
which allows for a shift of both the band centers and their widths.
For our choice of $n_b/n_a=2$, the amplitude in the first approximation $A_1 = 1.06$ 
accounts for the narrow band gaps observed  in the drawings.
\begin{figure} [htb]        
\includegraphics[width=7cm]{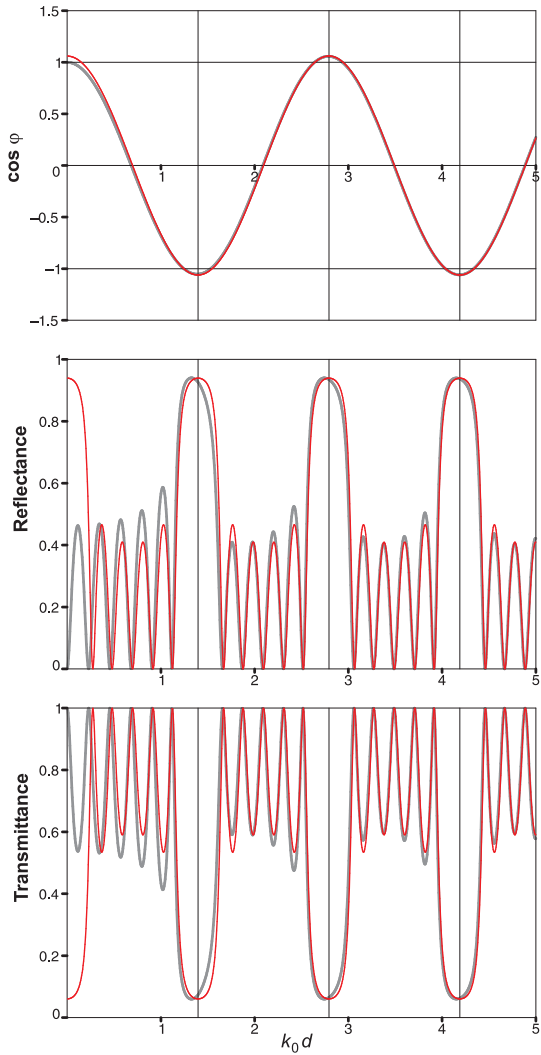} 
\caption{(colour online) Exact results (grey lines), see Eqs.~(18-20), 
{\it vs}. the first asymptotic approximation (red lines), see Eq.~(24), 
for the Bloch phase, reflectance, and transmittance,
for a sawtooth periodic structure with $n_a=1.5$, $n_b = 3$, 
and the number of periods $N=6$, in the case of normal incidence.  
The refractive indices outside the structure are $n_{\rm in}=1.0$ for $z<0$, 
and $n_{\rm ex}=1.0$ for $z>Nd$.}
\label{Fig2}
\end{figure} 
\begin{figure} [htb]        
\includegraphics[width=7cm]{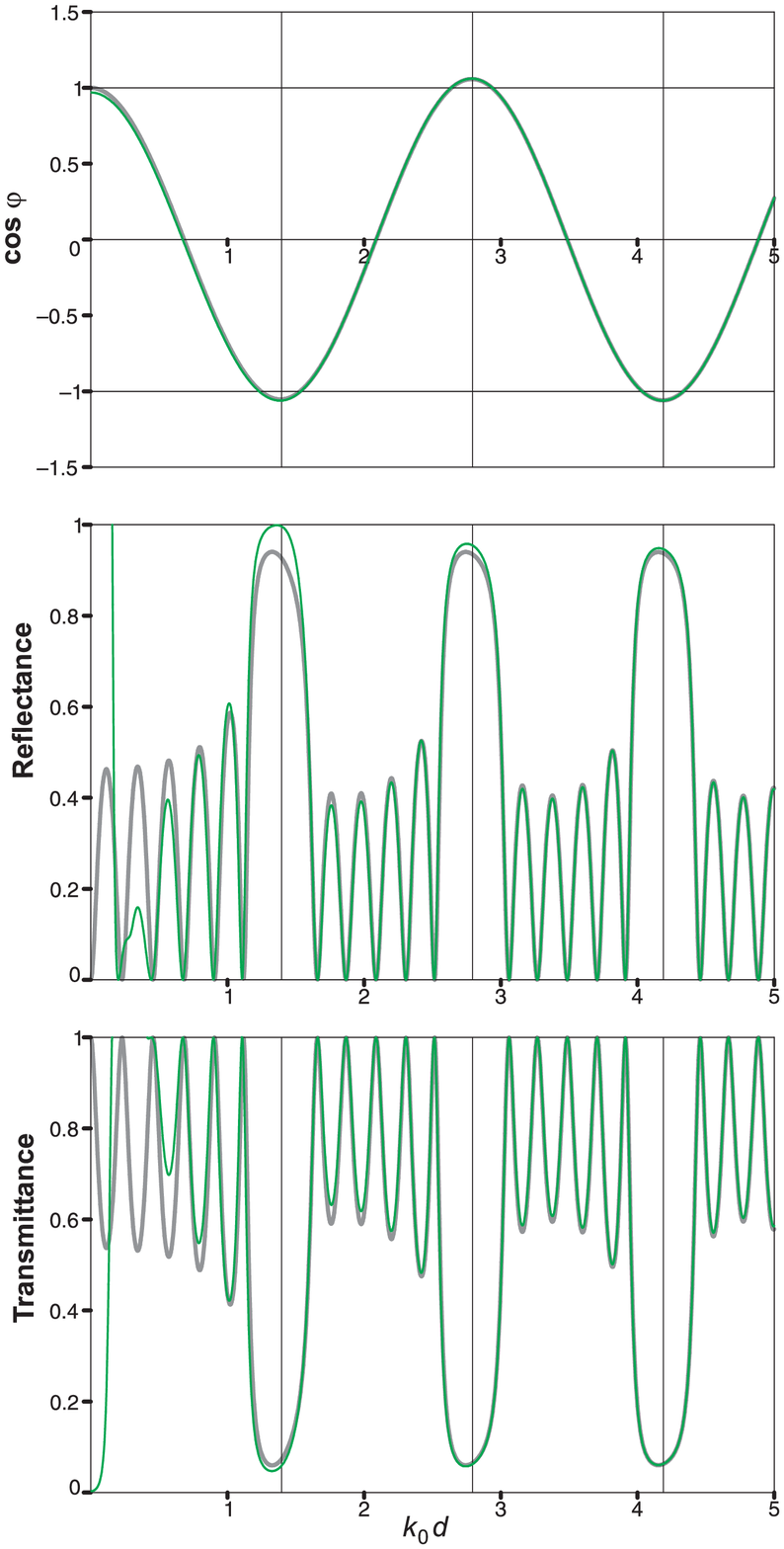}
\caption{(colour online) Exact results (grey lines), see Eqs.~(18-20), 
{\it vs}. the second asymptotic approximation (green lines), see Eq.~(25), 
for the Bloch phase, reflectance, and transmittance,
for a sawtooth periodic structure with parameters as in Fig.~2.}
\label{Fig3}
\end{figure} 
\begin{figure} [htb]        
\includegraphics[width=7cm]{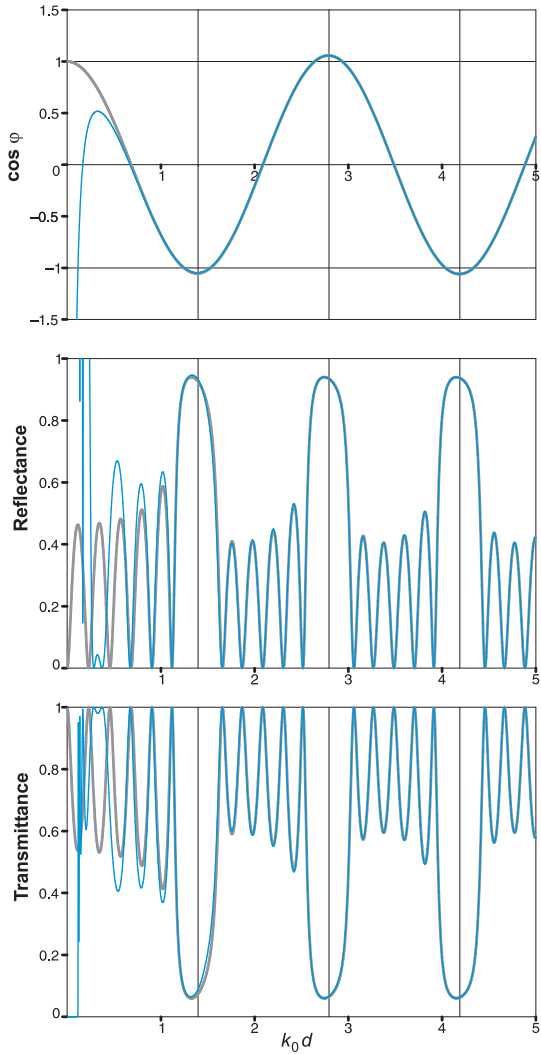}
\caption{(colour online) Exact results (grey lines), see Eqs.~(18-20), 
{\it vs}. the third asymptotic approximation (blue lines), see Eq.~(26), 
for the Bloch phase, reflectance, and transmittance,
for a sawtooth periodic structure with parameters as in Fig.~2.}
\label{Fig4}
\end{figure}

In Appendix B we also compare the exact results to  
approximations provided by binary photonic crystals 
(replacing the ramp by discrete steps).

\section{Bandgap Analysis}

Fig. \ref{Fig2} shows that the dispersion equation for $\cos \phi$ in the first approximation 
is already in quite good agreement with the exact results. 
We now discuss some of the predictions for the bandgaps that follow from it. 
The bandgap edges are defined by the condition $\cos\phi = (-1)^{q}$, where 
the index $q=1,2,\ldots$ numbers the bandgaps.
This condition applied to Eq.~(27) leads to expressions 
for the right $k_0^{r}$ and left $k_0^{l}$ bandedges in the first approximation: 
\begin{equation}                        
k_0^{r}\,d  = \frac{q\pi +\alpha}{n_{\rm av}}\,, \,\,\, k_0^{l}\,d = \frac{q\pi-\alpha}{n_{\rm av}}\,,\\
\end{equation}
where
\begin{equation}                        
\alpha  = \arccos \left(\frac{2}{\sqrt{n_a/n_b+n_b/n_a}}\right), \,\,\, n_{\rm av} =\frac{n_a+n_b}{2}.
\end{equation}
Therefore, the centers of the bandgaps $k_0^{c}$ and their widths $w$ 
are given  in the first approximation by 
\begin{equation}      									
k_0^{c}\,d = q\,\frac{\pi}{n_{\rm av}}, \quad wd = \frac{2\alpha}{n_{\rm av}}.
\end{equation}
This is qualitatively similar to the asymptotic results obtained by Rauh {\it et al}. \cite{Ra10}.  
In Fig.~5 we compare the accuracy of the three approximations for the band gap centers, 
and in Fig.~6 for their widths.  
As  one can see from Eqs.~(24)~-~(26), 
each approximation incorporates new corrections of orders $1/z_{a,b}^{\,2}$ 
and their squares respectively.
With the exception of the first bandgap 
(but that is inherent in use of an asymptotic series),
all three approximations provide one with quite accurate estimates
and the second one balances accuracy and simplicity. 
\begin{figure} [htb]                         
\includegraphics[width=8cm]{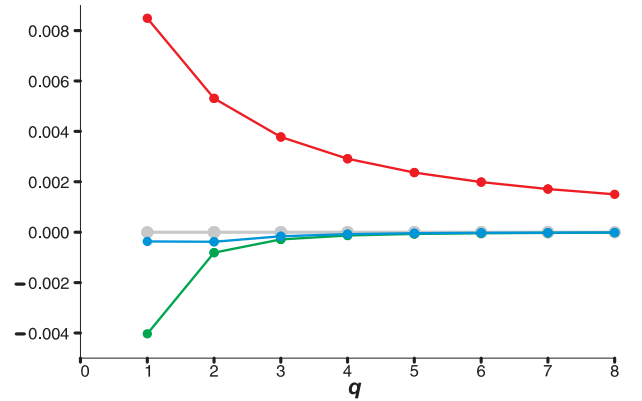}
\caption{(colour online) Dimensionless difference, 
$(k_{0,a}^c-k_0^c)\,d$, $a=1,2,3$, 
between exact results, $k_0^c\,d$, 
and a given asymptotic approximation, $k_{0,a}^c\,d$, 
for bandgap centers of a sawtooth periodic  array with refractive index parameters 
$n_a= 1.5$, $n_b = 3.0$  {\it vs}. the bandgap index $q$ for the first eight bandgaps. 
From top to bottom the lines are 
$(k_{0,1}^c-k_0^c)\,d$ 
(red), 
$(k_{0}^c-k_0^c)\,d$ 
(grey), 
$(k_{0,3}^c-k_0^c)\,d$ 
(blue), 
$(k_{0,2}^c-k_0^c)\,d$ 
(green).}
\label{Fig5saw}
\end{figure} 
\begin{figure} [htb]                         
\includegraphics[width=8cm]{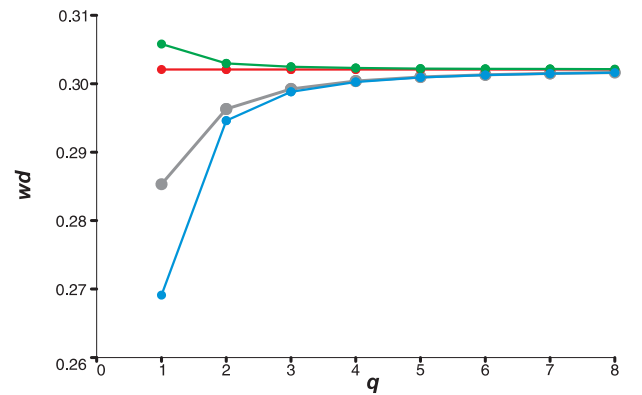}
\caption{(colour online) Dimensionless bandgap widths, $w\,d$,
for a sawtooth periodic  array with refractive index parameters 
$n_a= 1.5$, $n_b = 3.0$ 
{\it vs}. the bandgap index $q$ for the first eight bandgaps. 
From top to bottom the lines are 
the second asymptotic approximation (green), the first asymptotic approximation (red), 
exact result (grey), and the third asymptotic approximation (blue).} 
\label{Fig6saw}
\end{figure}

\section{Conclusion}

We have derived the exact dispersion equation, reflection, and transmission coefficients,
in terms of Bessel functions, for our one-dimensional photonic crystal with a sawtooth refractive index profile. 
Using the Hankel expansion, we worked out the first three asymptotic approximations to those coefficients.
Even the first approximation, which has a particularly simple analytic form, provides 
one with very reasonable estimates.  
Using the results of Appendix A, 
one can easily work out solutions for any monomial behaviour $(k_0 n(z))^2 \sim (z/L)^n$ 
of the permittitivity within a single cell of the periodic system.  They always involve Bessel functions 
$J_{\pm m}(z)$ of order $m=1/(n+2)$.   

\section{Acknowledgements}

We acknowledge support by FIS2011-24154 and 2009-SGR1289 (JM), 
and by NSERC discovery grant  RGPIN-3198 (DWLS).


\begin{figure} [htb]                         
\includegraphics[width=7cm]{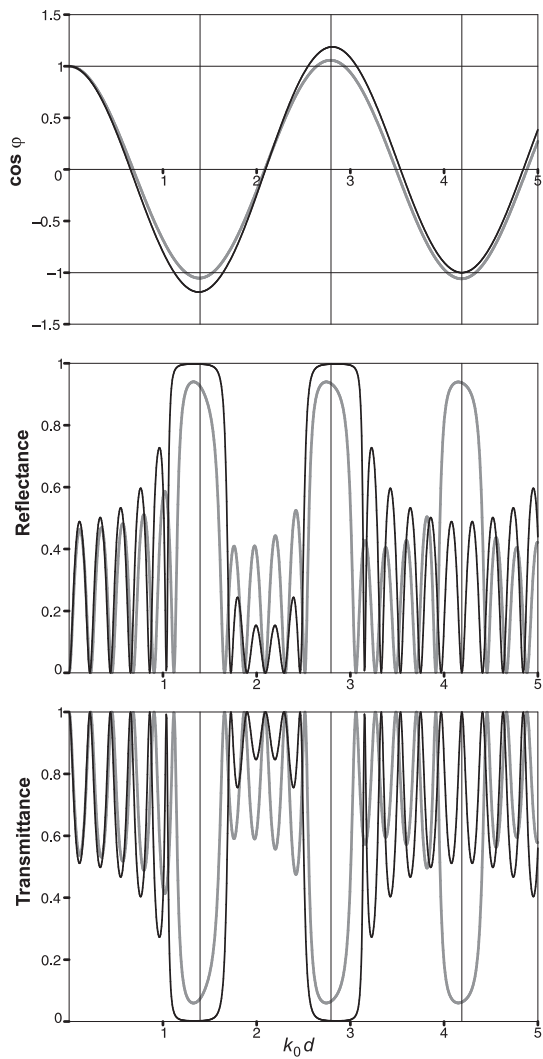}
\caption{(colour online) 
Exact results (grey lines) for a sawtooth periodic potential
with parameters as in Fig.~2 {\it vs}. binary approximation (black lines)
with $n_1=n_a$, $n_2=n_b$.} 
\label{Fig7saw}
\end{figure} 
\begin{figure} [htb]                         
\includegraphics[width=7cm]{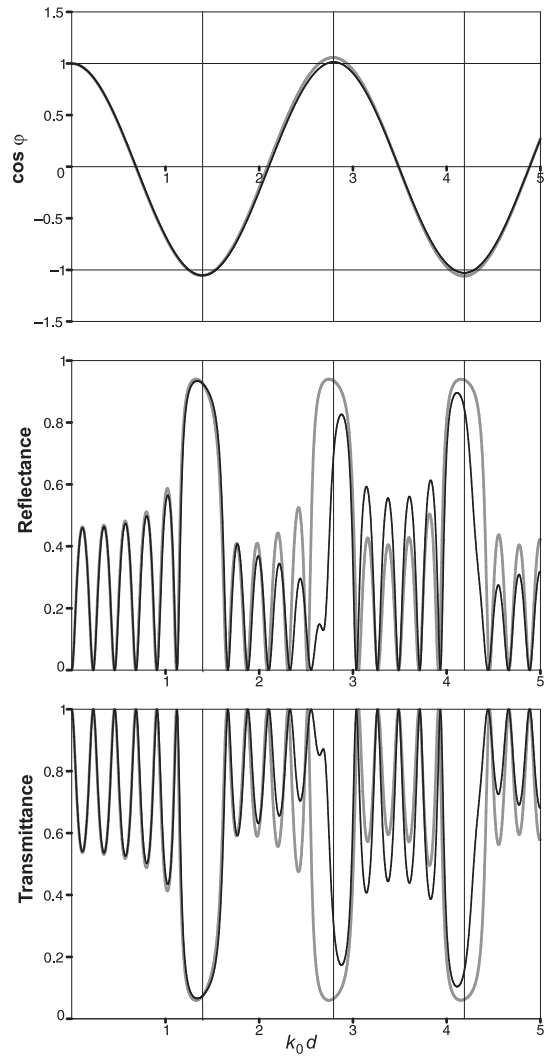}
\caption{(colour online) 
Like Fig.~7 but with $n_1 = (n_a + 3 n_b)/4$, $n_2 = (3 n_a + n_b)/4$.} 
\label{Fig8saw}
\end{figure} 

\appendix

\section{Solutions of the wave equation for power law potentials}

While discussing the WKB matching conditions at a classical turning point of order $n$, 
Schiff \cite{Schiff} points out that when $k^2(z) = (z/L)^n$,
an exact solution of the stationary state wave equation is given by 
\begin{equation}                        
0 = \psi^{\pr\pr}(z)  + k^2(z)\psi(z)\,, 
\label{eq:un01}
\end{equation} 
where
\begin{equation}                        
\begin{split}
\psi(z) & \equiv A(z) J_m(\xi)\,, \quad A(z) \equiv \sqrt{\xi(z)/k(z)}\,,\\ 
\xi(z) & \equiv  \int_0^z k(t) dt\,,  \quad \,\,\, m = \frac{1}{n+2}\,. 
\label{eq:un01a}
\end{split}
\end{equation} 
The task is to show that $J_m(\xi)$ is a generic Bessel function of order $m$, 
a result  given originally by Langer \cite{Lang37}.  The most used case 
is a linear turning point with $n=1$ $\Rightarrow$ $m=1/3$, where the solution coincides with the Airy function.  
The  general result makes it easy to work out the transfer matrix for a  1D photonic crystal whose
refractive index profile has any monomial ``sawtooth shape'' in each layer.
To begin, we note that for our assumed $k^2(z)$ 
\begin{equation}                        
\begin{split}
\xi(z) & = \frac{2L}{n+2} \Bigl(\frac{z}{L}\Bigr)^{(n+2)/2}, \\ 
k^\pr(z) & =  \frac{n}{2L} \Bigl(\frac{z}{L}\Bigr)^{(n-2)/2},\\ 
A^2(z) & =  \frac{\xi}{k} = 2 m z\,,\\
2 A(z) A^\pr(z) & = 2 m = {\rm const} \,,\\ 
A^3(z) A^{\pr\pr}(z) & =  -\left[A(z) A^\pr \right]^2 = -m^2. 
\label{eq:un03}
\end{split}
\end{equation} 
From here we work out $\psi^\pr$ and $\psi^{\pr\pr}$, and substitute into the wave equation. 
On functions of $z$, the prime means $d/dz$, but on the Bessel function it means $d/d\xi$. 
We have 
\begin{equation}                        
\begin{split}
\psi^\pr & = A^\pr(z) J_m(\xi) + A(z) k(z)  J_m^\pr(\xi),\\ 
\psi^{\pr\pr} & = A^{\pr\pr}(z) J_m(\xi)  + A(z) k^2(z)  J_m^{\pr\pr}(\xi)\\ 
              & + \Bigl[2 A^\pr(z) k(z) + A(z) k^\pr(z)\Bigr]  J^\pr_m(\xi) ~.  
\label{eq:un04}
\end{split}
\end{equation} 
Substituting into the wave equation and removing the factor $A k^2$, 
we find that the coefficient of $J^\pr_m(\xi)$ is 
\begin{eqnarray}                        
\frac{2 A^\pr k + Ak^\pr }{A k^2} &=& \frac{1}{\xi}~,  
\label{eq:un05}
\end{eqnarray} 
see Eq.~(\ref{eq:un08}) for details,
while the coefficient  of $J_m(\xi)$ is $1 + [A^{\pr\pr}/A k^2] $, where the 
variable  portion  is 
 \begin{eqnarray}               
\frac{A^{\pr\pr}}{A k^2} = 
\frac{A^3(z) A^{\pr\pr} }{A^4  k^2} &=& - \frac{ (A^{\pr})^2 A^2 }{A^4  k^2} = - \frac{m^2}{\xi^2}~.
\label{eq:un06}
\end{eqnarray} 
This verifies that the wave equation has been reduced to Bessel's equation for the fractional order $m$. 
 \begin{eqnarray} 			 
J^{\pr\pr}_m(\xi) + \frac{1}{\xi} J^\pr_m(\xi)  + \left[ 1 - \frac{m^2}{\xi^2} \right] J_m(\xi) &=&  0~. 
\label{eq:un07}
\end{eqnarray} 
The regular solution $J_m(\xi)$ will vanish at the origin when $m>0$, so the irregular solution $J_{-m}(\xi)$ is also required if a non-zero 
initial value is in order.
In Eq.~(\ref{eq:un05}) we used that
\begin{equation}                        
\begin{split}
A^2(z) =  \xi/k, & \quad 2 A A^\pr  = 1 - \xi \frac{k^\pr}{k^2}\,, \\  
\frac{2 A^\pr k + Ak^\pr }{A k^2} & = \frac{2 A A^\pr }{A^2 k} +  \frac{k^\pr}{k^2} = \frac{1}{\xi}\,.  
\label{eq:un08}
\end{split}
\end{equation} 

%

\section{Binary Approximations.}

We consider a binary photonic crystal composed of layers of constant refractive indices 
$n_1$, $n_2$, with corresponding thicknesses $d_1$, $d_2$. 
The elements of a binary $W$-matrix are well-known \cite{FPP} and given by
\begin{equation}
\begin{split}                           
W_{11} & = \cos(k_0n_1d_1)\cos(k_0n_2d_2)\\
& - \frac{n_2}{n_1}\sin(k_0n_1d_1)\sin(k_0n_2d_2),\\
\frac{W_{21}}{k_0} & = n_1\sin(k_0n_1d_1)\cos(k_0n_2d_2)\\
& + n_2\cos(k_0n_1d_1)\sin(k_0n_2d_2),\\
k_0W_{12} & = -\frac{1}{n_1}\sin(k_0n_1d_1)\cos(k_0n_2d_2) \\
& - \frac{1}{n_2}\cos(k_0n_1d_1)\sin(k_0n_2d_2),\\
W_{22} & =\cos(k_0n_1d_1)\cos(k_0n_2d_2)\\
& - \frac{n_1}{n_2}\sin(k_0n_1d_1)\sin(k_0n_2d_2).
\end{split}
\end{equation}

The simplest binary approximation to the sawtooth profile discussed above
is given by $n_1=n_a$, $n_2=n_b$, and $d_1 = d_2 = d/2$. 
Fig. \ref{Fig7saw} compares the exact results for the sawtooth potential 
with parameters as in Fig.~2 to that of the simplest binary approximation. 
As one can see there are some similarities, but the bandgap widths differ strongly;
for example, the third bandgap of the sawtooth potential disappears entirely.
A better binary approximation, see Fig. \ref{Fig8saw}, is obtained if one follows the recipe 
of Kalotas and Lee \cite{KL92}, taking the layer indices to be $n_1 = (n_a + 3 n_b)/4$, $n_2 = (3 n_a + n_b)/4$, 
and widths $d_1 = d_2 = d/2$. These values give the correct average wave number in each layer.


\end{document}